\documentclass[ALICE,manyauthors]{cernphprep}
\usepackage[comma,square,numbers,sort&compress]{natbib}
\usepackage{hyperref}

\begin{document}
\begin{titlepage}
\PHnumber{2012-065}
\PHdate{10 March 2012}
\title{Measurement of the cross section for electromagnetic dissociation with neutron emission in Pb--Pb Collisions at $\sqrt{s_{\rm NN}}$ = 2.76 TeV}
\ShortTitle{Electromagnetic dissociation with neutron emission}
\Collaboration{ALICE Collaboration%
         \thanks{See Appendix~\ref{app:collab} for the list of collaboration
                      members}}
\ShortAuthor{ALICE Collaboration}      
\begin{abstract}
The first measurement of neutron emission in electromagnetic dissociation of $^{208}$Pb nuclei at the LHC is presented. 
The measurement is performed using the neutron Zero Degree Calorimeters
of the ALICE experiment, which detect neutral particles close to beam rapidity.
The measured cross sections of single and mutual electromagnetic dissociation of Pb nuclei at $\sqrt{s_{\rm NN}}$ = 2.76 TeV with neutron emission are  
$\sigma_{\rm single\ EMD} = 187.4\pm0.2$~(stat.)~$^{+13.2} _{-11.2}$~(syst.)~b
and $\sigma_{\rm mutual\ EMD} = 5.7\pm0.1$~(stat.)~$\pm$0.4~(syst.)~b respectively.
The experimental results are compared to the predictions from a relativistic electromagnetic dissociation model.
\end{abstract}
\end{titlepage}
\setcounter{page}{2}
When two interacting nuclei collide at an impact 
parameter larger than the sum of the nuclear radii the interaction 
is purely electromagnetic.
The electromagnetic field of one of the two ions is experienced by the other ion as a flux of virtual photons.  
The equivalent photon method, proposed by Fermi~\cite{Fermi} in order to treat the moving electromagnetic field of a charged particle, was later extended by Weizs\"acker and Williams to collisions of ultra-relativistic electrons and protons with nuclei~\cite{Weizsacker, Williams}. 
As beam energy increases, the photon spectrum hardens and the flux is enhanced, 
due to the increase of the Lorentz contraction of the Coulomb field. Moreover the photon flux is proportional to $Z^2$, with $Z$ the charge number of the emitting nucleus.  Therefore the electromagnetic interactions become dominant in ultra-relativistic collisions of heavy-ions.
Two processes, the bound-free pair production and the electromagnetic dissociation (EMD), have attracted special attention in the last years, since they provide stringent limits on the beam lifetime in heavy-ion colliders~\cite{Bruce}. As predicted~\cite{RELDIS}, the excitation and subsequent decay of the Giant Dipole Resonance (GDR) via emission of one or two neutrons from colliding Pb nuclei occurs in $\sim$60\% of EMD events at the LHC. This can be exploited to measure the luminosity at heavy-ion colliders by detecting forward neutrons~\cite{White}.

This Letter reports the first measurement of the electromagnetic dissociation cross section of 
$^{208}$Pb nuclei at  $\sqrt{s_{\rm NN}}$ = 2.76 TeV via neutron emission, performed using the Zero Degree Calorimeters (ZDCs) of the ALICE experiment~\cite{ALICE} at the Large Hadron Collider (LHC).  The ZDCs are ideally suited to tag EMD interactions, since the resulting neutrons from the GDR decay are emitted very close to beam rapidity and are the most abundant particles produced in these processes. The data were collected using the neutron ZDCs (ZNA and ZNC), located 
114 m away from the Interaction Point (IP) at the so-called A and C sides of the ALICE detector.
Each ZN is placed at zero degree  with respect to the LHC beam axis and is used to detect neutral particles at pseudo--rapidities $|\eta| > 8.7$.  
For the present analysis two small forward electromagnetic calorimeters (ZEM1 and ZEM2), placed on the A side at 7.35 m from the IP 
($4.8 \le \eta \le 5.7$), are also used to tag hadronic interactions.

The experimental results are presented and compared to theoretical predictions of the Relativistic ELectromagnetic DISsociation (RELDIS) model~\cite{RELDIS}, which is designed to describe electromagnetic interactions between  ultra-relativistic nuclei including single and double virtual photon absorption, excitation of giant resonances, intra--nuclear cascades of produced hadrons and statistical decay of excited residual nuclei. 
Above the GDR region photon-induced reactions become more complicated leading to multiple ($>$3) emission of neutrons~\cite{Lepretre82}.  
RELDIS accurately reproduces this experimental observation and also predicts further increase of  the mean number of neutrons and of the width of their multiplicity distribution as photon energy increases~\cite{Pshenichnov99}. 
Calculations based on this model provide a good description of neutron emission in electromagnetic dissociation of Pb ions at the CERN SPS~\cite{CERN} and of Au  ions at the Relativistic Heavy Ion Collider (RHIC)~\cite{RHIC}.
\begin{figure}[hbt]
\begin{center}
\includegraphics[width=0.6\linewidth]{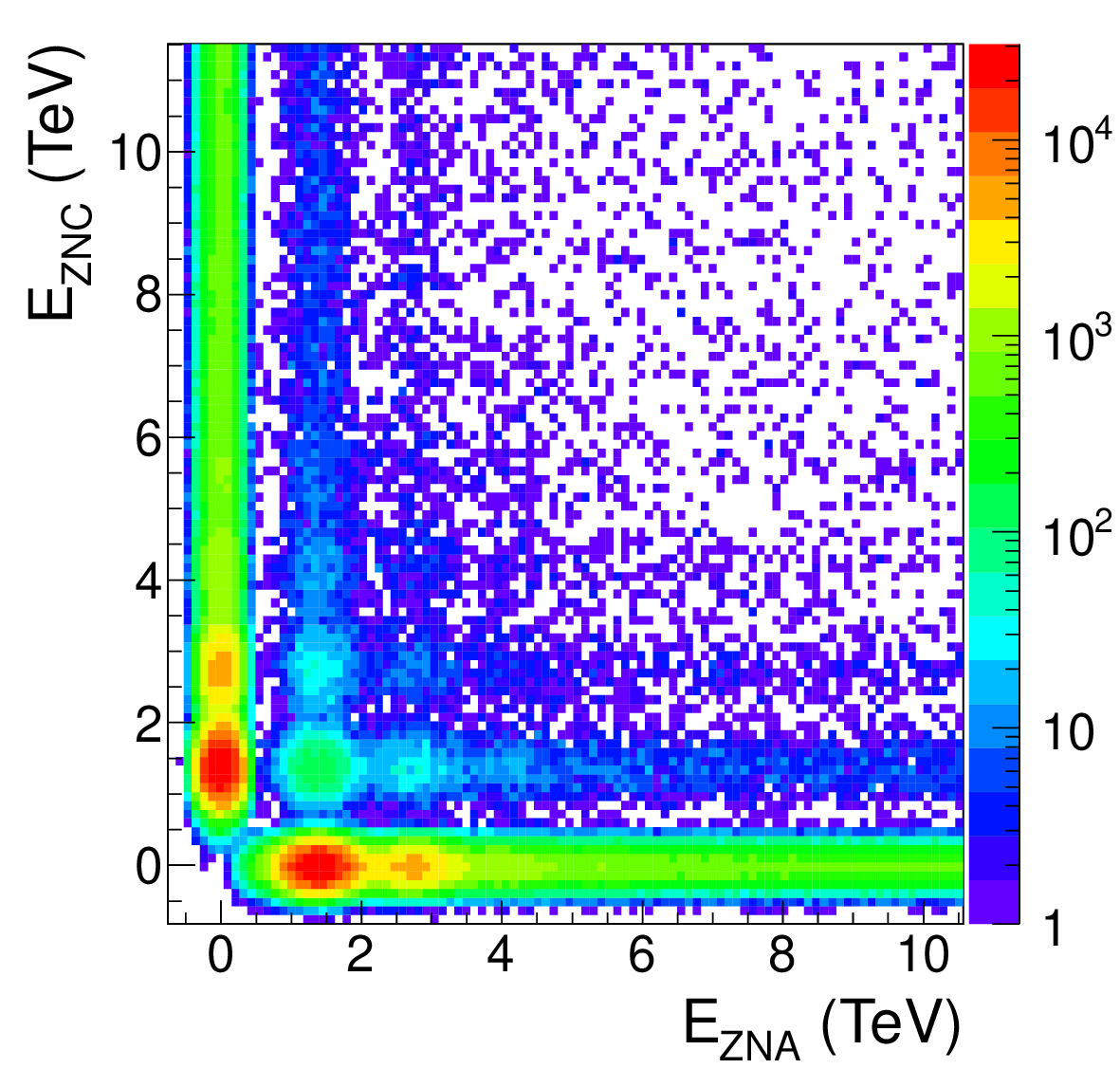}
\caption{\label{fig:fig1}
Energy deposition in ZNC versus ZNA for single EMD plus hadronic events. The 1n signal is at 1.38 TeV. The events where at least 1n is detected by both ZNs are associated to mutual EMD and hadronic processes.}
\end{center}
\end{figure}

During the $\sqrt{s_{\rm NN}}$ = 2.76 TeV Pb--Pb data taking in 2010, an EMD run was performed.
In this dedicated run only the ZDCs and ZEM were read--out.
The trigger was set to tag neutrons emitted in EMD as well as hadronic interactions (see Figure~\ref{fig:fig1}),
requiring a minimum energy deposit in at least one of the two ZNs ($\sim3\times10^6$ events were collected).  
The energy thresholds were $\sim$450 GeV for ZNA and $\sim$500 GeV for ZNC and were placed approximately three standard deviations below the energy deposition of a 1.38 TeV neutron. 
The depletion of events in the region where the ZNA and ZNC energy deposition is close to 0 TeV is related to the (ZNA OR ZNC) trigger onset.

Following a common convention we define as single EMD a process where at least one neutron (1n) is emitted by a given Pb nucleus disregarding the fate of the other nucleus.   
Mutual EMD events, where at least 1n is emitted by both Pb nuclei,  
and hadronic events were selected offline requiring an energy deposit above the energy threshold in both ZNs.

In the 2010 Pb--Pb run ZNs were used as the ALICE luminometer, providing different logical combinations of signals (ZDC triggers). 
In particular during a beam separation van der Meer (vdM) scan~\cite{vdM}, 
 a cross section 
$\sigma_{\rm {ZNA \: OR \: ZNC}}^{\rm {vdM}}$ = 371.4 $\pm$ 0.6 (stat.) $^{+24} _{-19}$ (syst.) b was measured for the (ZNA OR ZNC) trigger, tagging single EMD plus hadronic interactions.
The systematic error of $-5.2\%+6.4\%$ can be decomposed as follows:  $4.3\%$ uncertainty coming from the vdM scan analysis~\cite{vdMA}, dominated by the calibration of the distance scale during the scan; $-3\%+4.7\%$ uncertainty coming from the measurement of the beam intensity, dominated by the beam current transformers scale~\cite{Alici} and by the non-colliding (ghost) charge fraction in the LHC beams~\cite{Jeff,BCNWG}. The beam-gas contribution ($\sim 2.5\%$) is subtracted.

\begin{figure}[hbt]
\begin{center}
\includegraphics[width=0.6\linewidth]{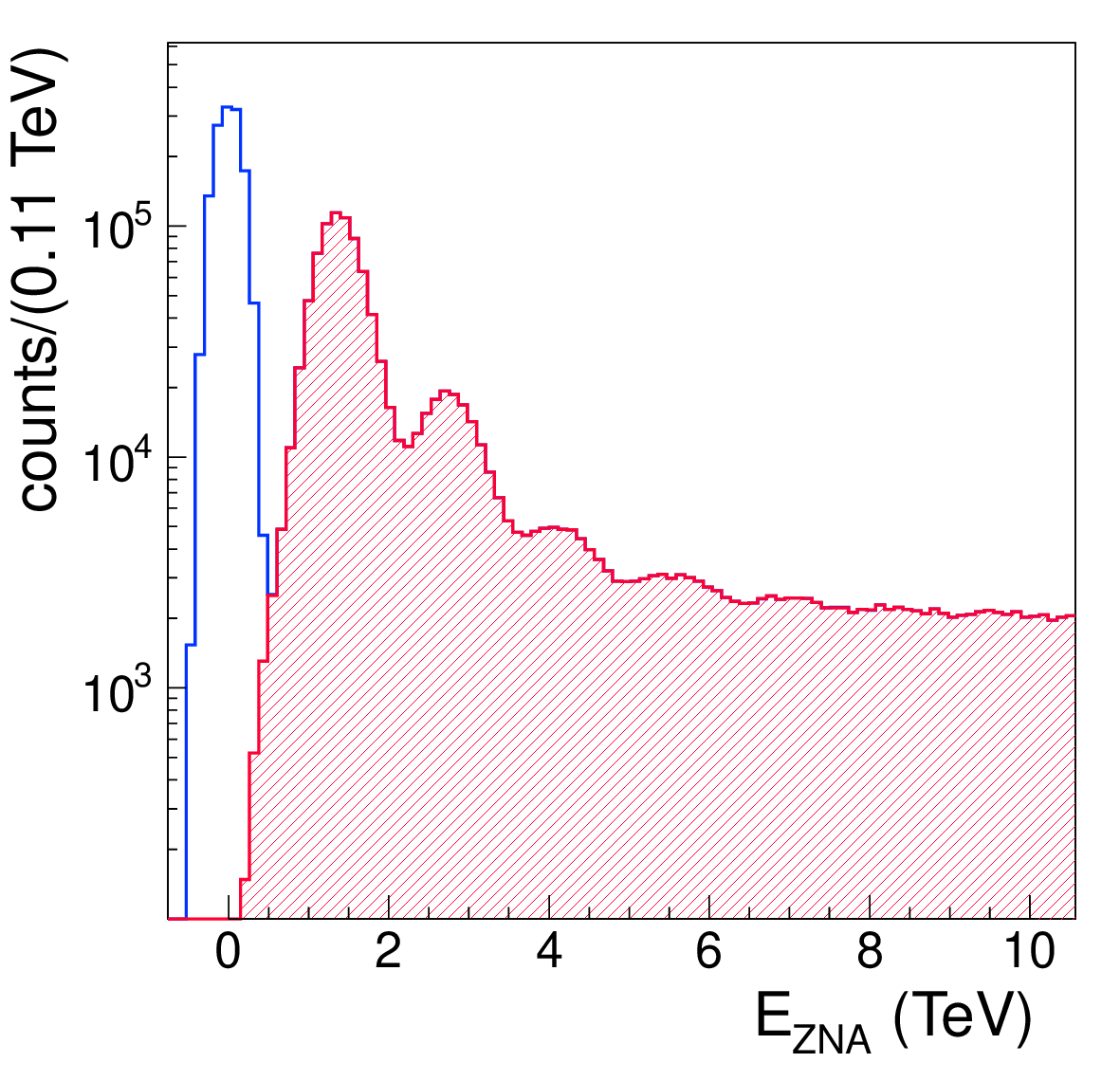}
\caption{\label{fig:fig2}
ZNA energy spectrum requiring signal over threshold in ZNA or ZNC (not filled area) superimposed to ZNA energy spectrum requiring signal in ZNA (shaded area). The first peak centered at $E=0$ corresponds to pedestal events,where no signal from neutron emission is detected by the ZNA.}
\end{center}
\end{figure}

The energy spectrum for the ZNA is shown in Figure~\ref{fig:fig2}, for events 
in which there is a signal in at least one of the two ZNs (not filled area) or  for events in which ZNA is fired (shaded area). The selection of events with signal in ZNA is performed offline using the timing information provided by a TDC (Time to Digital Converter). This provide a sharper cut with respect to a selection based on energy deposit. 
In the first case a pedestal peak centered at $E=0$ is visible, which corresponds to events where no signal is detected by the ZNA and the trigger is fired by the ZNC. As can be inferred in Figure~\ref{fig:fig2} the TDC selection rejects only events in the pedestal. The width of the pedestal peak is related to the noise of electronic modules. 
In the energy spectrum a pronounced 1n peak at 1.38 TeV is present, but also 2n, 3n, 4n... peaks are clearly identified.  
The requirement of a signal in the TDC for the ZNA and the ZNC, respectively,  allows to calculate two different estimates of the number of events from single EMD plus hadronic processes. The average of the two results is then calculated (the difference between the response of the ZNA and the ZNC is about 0.1\%). The contamination from beam-residual gas interactions, estimated via the observed rates with circulating beams, before they are brought into collisions, 
is of the order of 2.5\% and is corrected for.

\begin{figure}[hbt]
\begin{center}
\includegraphics[width=0.6\linewidth]{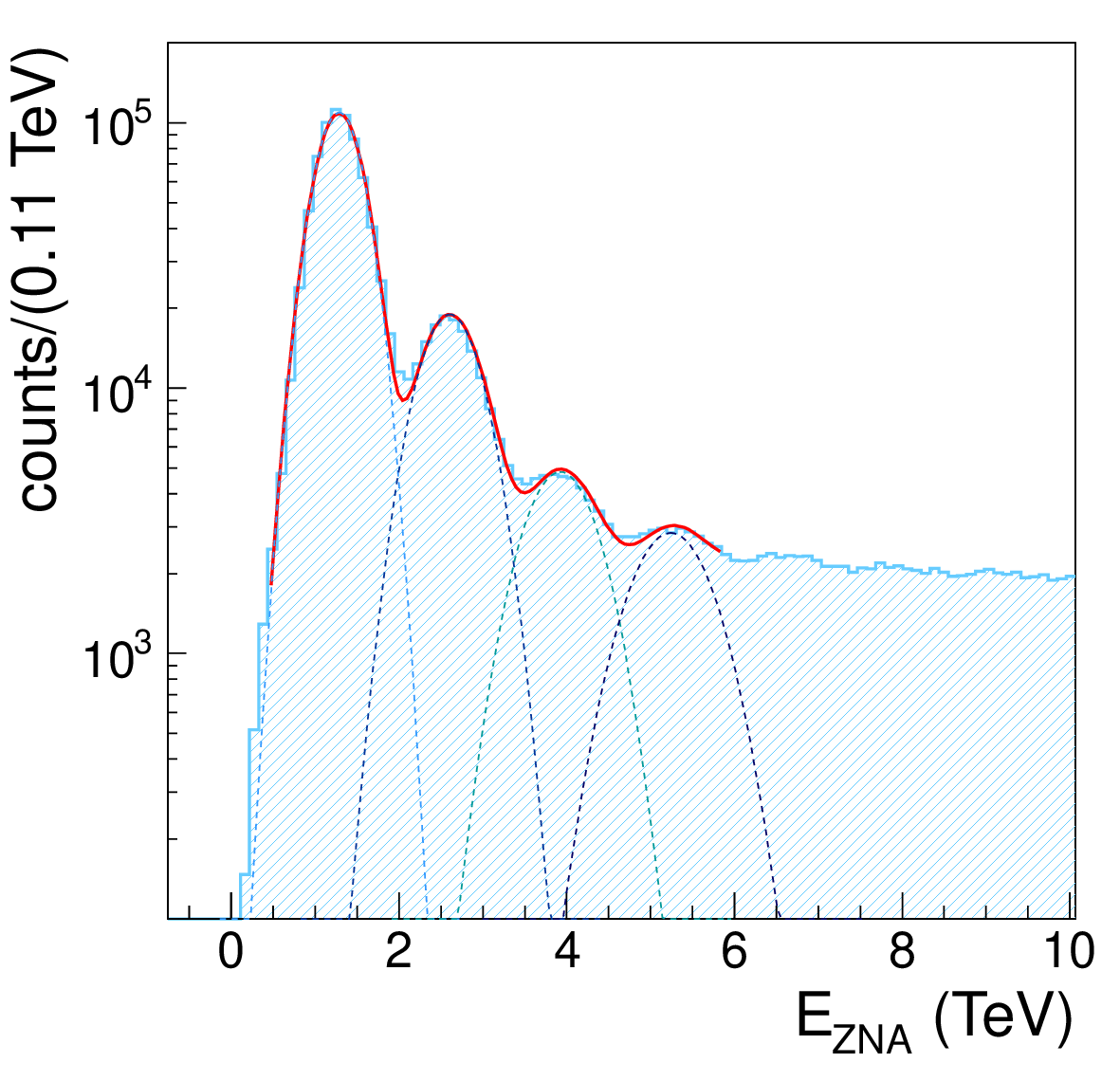}
\caption{\label{fig:fig3}
ZNA energy spectrum requiring signal over threshold in ZNA but not in ZNC, rejecting thus neutron emission on the opposite side. The dashed lines represent the single fits of the different peaks (1n, 2n,...), while the continuous line is the sum of all the contributions.}
\end{center}
\end{figure}

A second event selection requires a signal in one of the ZNs, but not in the other one. In this way hadronic events, which mostly lead to disintegration of both colliding nuclei, are rejected. 
In this case the mutual EMD events are also removed from the spectrum and therefore the selected process is the single EMD minus the mutual EMD.
The energy spectrum is shown in Figure~\ref{fig:fig3} together with the fit obtained by summing four Gaussians. 
The curve for the 1n peak has three free parameters, while the following Gaussians describing the $\it i^{th}$ peak have a constraint both on the 
mean value 
$\mu_{{\it i}{\rm n}}$~($\mu_{{\it i}{\rm n}} = \it i \times \mu_{\rm 1n}$, 
where $\mu_{{\it i}{\rm n}}$ is the mean value for $\it i^{th}$ neutron peak)  
and on the width $\sigma_{{\it i}{\rm n}}$~($\sigma_{{\it i}{\rm n}} = \sqrt{\it i \times (\sigma_{\rm 1n}^2 - \sigma_{\rm ped}^2) + \sigma_{\rm ped}^2}$, where $\sigma_{{\it i}{\rm n}}$ is the width of the $\it i^{th}$ neutron peak and $\sigma_{\rm ped}$ is the width of the pedestal peak).  
The relative energy resolution $\sigma_{1{\rm n}} / \mu_{1{\rm n}}$ of the 1n peak at 1.38 TeV is 21\% for the ZNA and 20\% for the ZNC, in agreement with expectations from beam tests at the CERN SPS~\cite{ZDC} 
extrapolated to LHC energies using Monte Carlo, which takes into account the different operating conditions.  
Similarly to the previous analysis we made the average of the ZNA and the ZNC cross sections, which difference is about 0.2\%. 

\begin{table}[hbt]
\centering
\caption{Cross sections (barn) for $\sqrt{s_{\rm NN}}$ = 2.76 TeV Pb--Pb interactions (systematic errors are dominated by the vdM cross section errors).
Theoretical uncertainties are  systematic and related to uncertainties in the total photoabsorption cross sections on Pb.}
\begin{tabular}{@{} c|c|c @{}} 
\hline
\hline
Physical Process &Data  &RELDIS\\\hline
single EMD + & 194.8$\pm$0.3 stat. $^{+13.6} _{-11.5}$ syst. & 192.9$\pm$9.2 \\
hadronic & & \\ \hline
single EMD - & 181.3$\pm$0.3 stat. $^{+12.8} _{-10.9}$ syst. & 179.7$\pm$9.2 \\
mutual EMD & & \\ \hline \hline
mutual EMD & 5.7$\pm$0.1 stat. $\pm$0.4 syst. & 5.5$\pm$0.6 \\ \hline  hadronic & 7.7$\pm$0.1 stat. $^{+0.6} _{-0.5}$ syst. & 7.7$\pm$0.4 \\ \hline \hline
single EMD & 187.4$\pm$0.2 stat. $^{+13.2} _{-11.2}$ syst. & 185.2$\pm$9.2 \\ \hline
\hline
\end{tabular}
\label{tab:cross}
\end{table}

The cross sections, listed in Table~\ref{tab:cross} (first two rows), are calculated using the (ZNA OR ZNC) cross section measured during the vdM scan:  $\sigma_{\rm proc} = \sigma_{\rm {ZNA \: OR \: ZNC}}^{\rm {vdM}} \times N_{\rm proc} / N_{\rm {ZNA \: OR \: ZNC}}$, where $N_{\rm proc}$ is the number of events in the sample of the selected process and $N_{\rm {ZNA \: OR \: ZNC}}$ is the number of events collected with the same trigger as used to determine $\sigma_{\rm {ZNA \: OR  \: ZNC}}^{\rm {vdM}}$.
The calculated values are corrected for the ZN detection probability 
(98.7\% $\pm$ 0.04\%(stat.) $\pm$ 0.1\%(syst.)), 
estimated from a Monte Carlo simulation using RELDIS as event generator.
The systematic errors, dominated by the uncertainties of the cross sections measured during the vdM scan, take also into account the difference between the response of the ZNA and the ZNC (0.1-0.2\%) and the uncertainty due to the estimate of beam-gas background ($\sim 1\%$).
The centering of ZN calorimeters on the neutron spot was assured by the 
measurement of the centroid position, thanks to their transverse segmentation 
in four towers.

The predictions of the RELDIS model for $\sqrt{s_{\rm NN}}$ = 2.76 TeV Pb--Pb EMD interactions are also shown in  Table~\ref{tab:cross}.  The agreement between data and model predictions is remarkable.
In calculations of the EMD cross sections various approximations of the total photoabsorption cross sections on lead are used, leading to 5\% uncertainties in the predicted values~\cite{RELDIS}.  These errors include the difference between RELDIS and other theoretical predictions~\cite{Baltz}.

\begin{figure}[hbt]
\begin{center}
\includegraphics[width=0.6\linewidth]{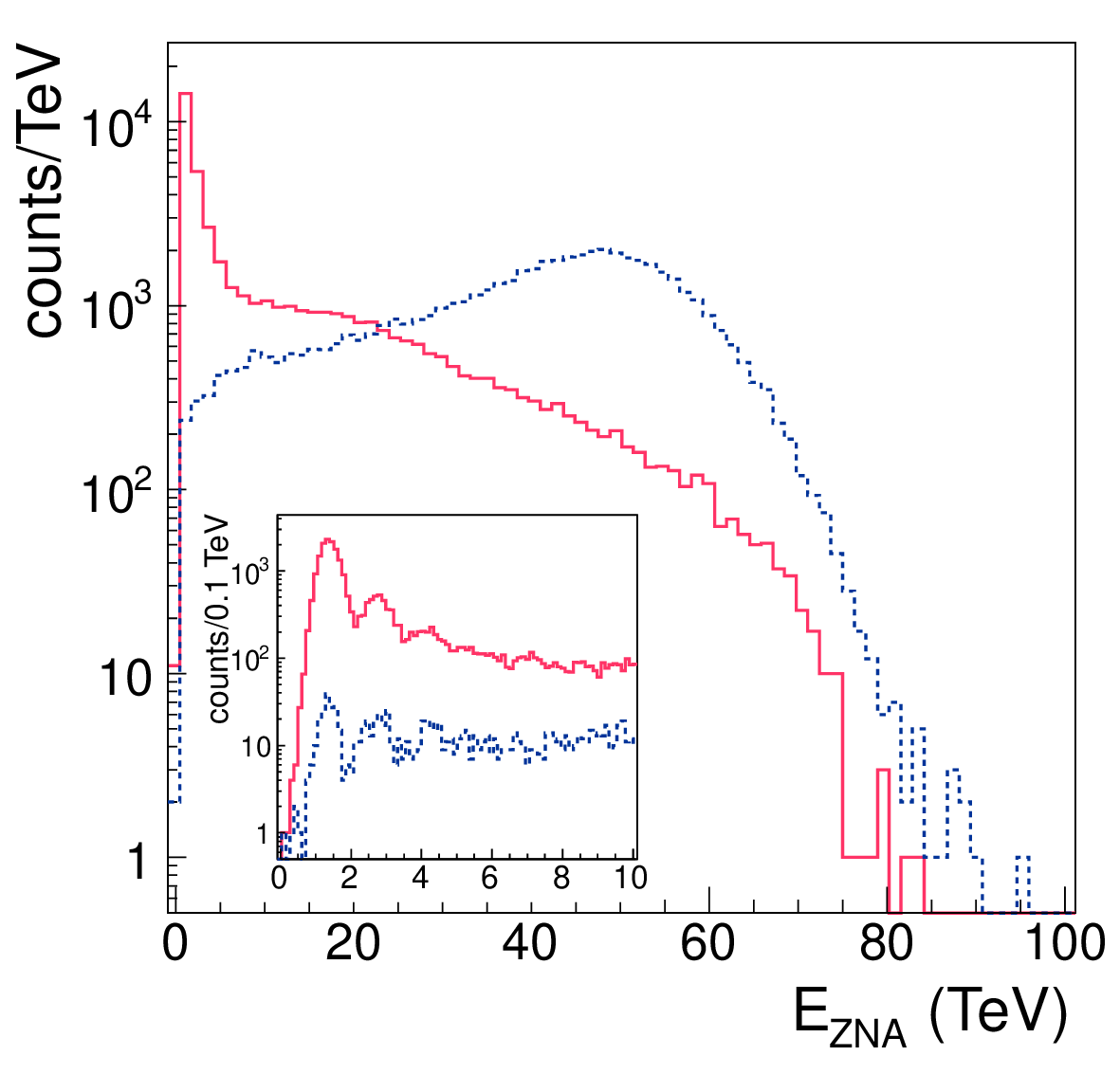}
\caption{\label{fig:fig4}
ZNA energy spectrum for mutual EMD (no signal in any ZEM, continuous line) and hadronic (a signal in at least one of the two ZEMs, dashed line) event selection. The insert shows an expanded view of the low energy region.}
\end{center}
\end{figure}

A third event selection is performed to select mutual EMD and hadronic events requiring a minimum energy deposition in both ZNs. This selection rejects all beam-gas contributions. To disentangle the mutual EMD and the hadronic processes, the ZEMs are used to select events with no signal in any ZEM or a signal in at least one of the two ZEMs, respectively. The energy threshold for each ZEM is about 10 GeV. 
Figure~\ref{fig:fig4} shows the ZNA energy spectrum for the mutual EMD (continuous line) and hadronic (dashed line) event selection.
\begin{figure}[hbt]
\begin{center}
\includegraphics[width=0.6\linewidth]{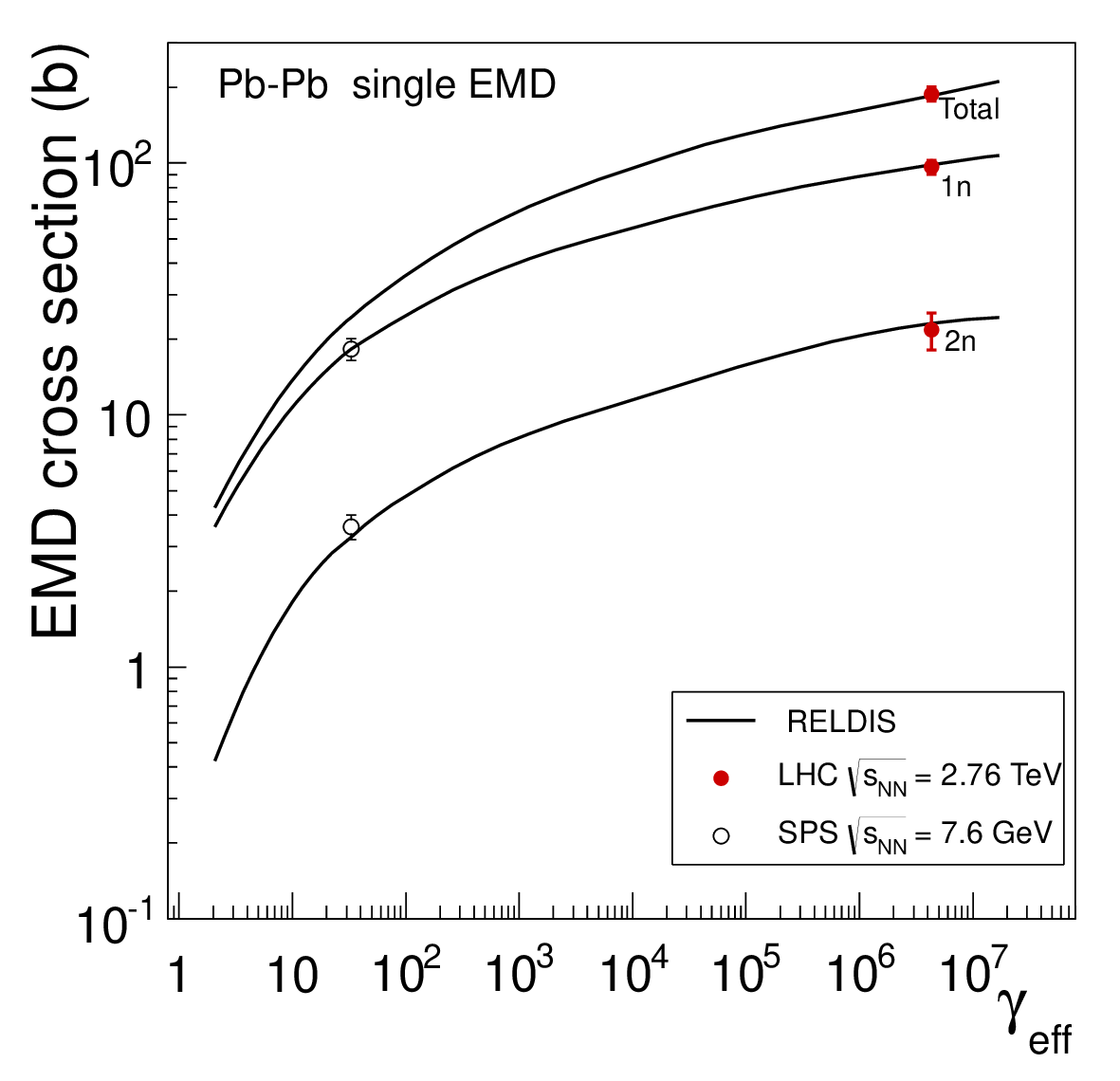}
\caption{\label{fig:fig5}
Total single EMD cross sections and partial EMD cross sections for 
emission of one and two neutrons as a function of the effective Lorentz factor $\gamma_{\rm eff}$. 
The closed symbols are our data, while the open symbols represent the results obtained at CERN SPS~\cite{CERN} at 30 GeV. 
The RELDIS predictions~\cite{CERN} for total, 1n and 2n EMD cross sections are shown as solid lines.}
\end{center}
\end{figure}
The cross sections for the mutual EMD and hadronic processes are calculated, as in the previous analysis, using the vdM (ZNA OR ZNC) cross section. 
The ZEM trigger efficiencies for the mutual EMD event selection, i.e. the fraction of mutual EMD events with no signal in any ZEM, is  
96.0\% $\pm$ 0.1\%(stat.) $\pm$ 0.6\%(syst.), 
evaluated from simulation using RELDIS as event generator.
The ZEM trigger efficiencies for the hadronic event selection, i.e. the
the fraction of hadronic events with a signal in at least one of the two ZEMs, is 92.4\% $\pm$ 0.3\%(stat.) $\pm$ 1.0\%(syst.), 
estimated using HIJING~\cite{higing} as event generator, combined 
with a simple fragmentation model~\cite{TDR}.
Since the two event selections are mutually exclusive,
the contamination of mutual EMD events in the hadronic sample and of hadronic events in the mutual EMD sample are $\sim 4\%$ and $\sim 7.6\%$ respectively.

The raw cross sections ($\sigma_{mEMD,\, raw}$, $\sigma_{hadr,\, raw}$) 
and the ZEM trigger efficiencies ($\epsilon_{mEMD}$, $\epsilon_{hadr}$)
for the two processes are inserted in a system of equations in two variables, 
where the unknowns are the true mutual EMD and the 
true hadronic cross sections ($\sigma_{mEMD,\, true}$, $\sigma_{hadr,\, true}$), 
respectively. 
\[\begin{cases}
\sigma_{mEMD,\, raw} &
=\epsilon_{mEMD}\cdot\sigma_{mEMD,\, true}+\left(1-\epsilon_{hadr}\right)\cdot\sigma_{hadr,\, true}\\
\sigma_{hadr,\, raw} &
=\left(1-\epsilon_{mEMD}\right)\cdot\sigma_{mEMD,\, true}+\epsilon_{hadr}\cdot\sigma_{hadr,\, true}
\end{cases}\]
The extracted values are corrected for the estimated ZN detection probability for mutual EMD (95.7\% $\pm$ 0.07\%(stat.) $\pm$ 0.5\%(syst.)) and for hadronic  (97.0\% $\pm$ 0.2\%(stat.) $\pm$ 3\%(syst.)) events.
The mutual EMD cross section is also corrected for background from accidental coincidences between uncorrelated single EMD interactions ($\sim 10\%$).
The final cross section results are summarized and compared to the RELDIS predictions in  Table~\ref{tab:cross} (third and fourth rows).  

The single EMD cross section  listed in Table~\ref{tab:cross} (last row) is estimated from previous measurements, making an average of the (single EMD + hadronic) $-$ hadronic and the (single EMD $-$ mutual EMD) + mutual EMD cross sections.

For the single EMD minus mutual EMD event selection the measured fractions of 1n, 2n and 3n events with respect to the total number of events is estimated (Table~\ref{tab:ratio}).
The table contains also the relevant expectations for the ratios based on the calculations with the RELDIS model.
The 1n and 2n emission channels give the main contribution (63\%), confirming that EMD processes proceed predominantly via GDR excitation and subsequent decay by neutron emission.
According to RELDIS, 3n emission is mostly induced by energetic ($>$40 MeV) equivalent photons and frequently accompanied by emission of protons and pions.
The measured 1n and 2n yields are much closer to RELDIS predictions compared to the 3n yields. This can be explained by the fact that RELDIS was already tuned  by comparison with 1n and 2n  data on photoabsorption on lead~\cite{RELDIS} and on EMD of 30 A GeV lead nuclei~\cite{CERN}. Unfortunately, the data on neutron emission induced by photons above 140 MeV are absent, while according to RELDIS almost half of 3n events is due to such energetic photons. 
In EMD calculations the native photonuclear reaction model of RELDIS can be replaced by the GNASH code~\cite{Chadwick95}, thus providing slightly different results for 1n and 2n yields.
On the basis of this difference the theoretical uncertainties listed in Table~\ref{tab:ratio} are estimated.

\begin{table}[hbt]
\centering
\caption{Neutron emission fractions for single EMD minus mutual EMD process in $\sqrt{s_{\rm NN}}$ = 2.76 TeV Pb--Pb interactions.
Theoretical uncertainties are systematic and related to the divergence of predictions of various photonuclear reaction models.}
\begin{tabular}{@{} c|c|c @{}} 
\hline
\hline
Ratio & Data(\%)  & RELDIS(\%)\\\hline
1n/N$_{\rm tot}$ & 51.5$\pm$0.4 stat. $\pm$0.2 syst. &54.2$\pm$2.4 \\ 
2n/N$_{\rm tot}$ & 11.6$\pm$0.3 stat. $\pm$0.5 syst. &12.7$\pm$0.8 \\ 
3n/N$_{\rm tot}$ & 3.6$\pm$0.2 stat. $\pm$0.2 syst. &5.4$\pm$0.7 \\ 
2n/1n & 22.5$\pm$0.5 stat. $\pm$0.9 syst.&23.5$\pm$2.5 \\ 
\hline
\hline
\end{tabular}
\label{tab:ratio}
\end{table}
Our 2n to 1n ratio of (22.5$\pm$0.5$\pm$0.9)\% in single EMD can be compared to the value of (19.7$\pm$2.9)\% reported for Pb--Pb collisions at 30 A GeV at the CERN SPS~\cite{CERN}.  
As predicted by RELDIS, the observed weak increase (around one standard deviation) of the 2n to 1n ratio with collision energy is due to additional 2n events produced by more energetic equivalent photons at the LHC.

Finally, Figure~\ref{fig:fig5} presents total and partial EMD cross sections 
for emission of one and two neutrons measured by ALICE compared to CERN SPS data~\cite{CERN}. 
The results of the RELDIS model are also shown for 
a wide range of the projectile effective Lorentz-factor $\gamma_{\rm eff}$ calculated in the rest frame of the collision partner.  As seen, both data 
sets are successfully described by the model despite of 
six orders-of-magnitude span of $\gamma_{\rm eff}$.
A direct comparison to RHIC results is not straightforward since the structure of the involved nuclei is different.  Since $^{208}$Pb is a double magic nucleus, while $^{197}$Au is not, the GDR position, its width as well as the neutron emission thresholds differ in such nuclei.


In summary a first measurement of electromagnetic dissociation in $\sqrt{s_{\rm NN}}$ = 2.76 TeV Pb--Pb  collisions was performed at the LHC by detection of the emitted neutrons with the ALICE ZDCs. The measurement tests the theoretical predictions used for estimations of beam losses. The RELDIS model predictions are in a very good agreement with our experimental results. 
The measurements reported here establish experimentally the EMD cross section scale for the first time at LHC energy. We finally note that the ALICE ZDC detectors, calibrated through these results, provide the possibility of a direct absolute measurement of the LHC luminosity in Pb--Pb collisions.

%

\bibliographystyle{apsrev4-1}
\bibliography{emdPbPb}{}

\begin{thebibliography}{38}%
\makeatletter
\providecommand \@ifxundefined [1]{%
 \@ifx{#1\undefined}
}%
\providecommand \@ifnum [1]{%
 \ifnum #1\expandafter \@firstoftwo
 \else \expandafter \@secondoftwo
 \fi
}%
\providecommand \@ifx [1]{%
 \ifx #1\expandafter \@firstoftwo
 \else \expandafter \@secondoftwo
 \fi
}%
\providecommand \natexlab [1]{#1}%
\providecommand \enquote  [1]{``#1''}%
\providecommand \bibnamefont  [1]{#1}%
\providecommand \bibfnamefont [1]{#1}%
\providecommand \citenamefont [1]{#1}%
\providecommand \href@noop [0]{\@secondoftwo}%
\providecommand \href [0]{\begingroup \@sanitize@url \@href}%
\providecommand \@href[1]{\@@startlink{#1}\@@href}%
\providecommand \@@href[1]{\endgroup#1\@@endlink}%
\providecommand \@sanitize@url [0]{\catcode `\\12\catcode `\$12\catcode
  `\&12\catcode `\#12\catcode `\^12\catcode `\_12\catcode `\%12\relax}%
\providecommand \@@startlink[1]{}%
\providecommand \@@endlink[0]{}%
\providecommand \url  [0]{\begingroup\@sanitize@url \@url }%
\providecommand \@url [1]{\endgroup\@href {#1}{\urlprefix }}%
\providecommand \urlprefix  [0]{URL }%
\providecommand \Eprint [0]{\href }%
\providecommand \doibase [0]{http://dx.doi.org/}%
\providecommand \selectlanguage [0]{\@gobble}%
\providecommand \bibinfo  [0]{\@secondoftwo}%
\providecommand \bibfield  [0]{\@secondoftwo}%
\providecommand \translation [1]{[#1]}%
\providecommand \BibitemOpen [0]{}%
\providecommand \bibitemStop [0]{}%
\providecommand \bibitemNoStop [0]{.\EOS\space}%
\providecommand \EOS [0]{\spacefactor3000\relax}%
\providecommand \BibitemShut  [1]{\csname bibitem#1\endcsname}%
\let\auto@bib@innerbib\@empty
\bibitem{Fermi}
E. Fermi, Z. Phys. {\bf29}, 315 (1924).

\bibitem{Weizsacker}
C.F. von Weizs\"acker, Z. Phys. {\bf 88}, 612 (1934). 

\bibitem{Williams}
E.J. Williams, Phys. Rev. {\bf 45}, 729 (1934).

\bibitem{Bruce}
R. Bruce et al., Phys. Rev. ST Accel. Beams {\bf 12}, 071002 (2009).

\bibitem{RELDIS}
I.A. Pshenichnov et al., Phys. Rev. C {\bf 64}, 024903 (2001); I.A. Pshenichnov, Phys. Part. Nucl. {\bf 42}, 215 (2011).

\bibitem{White}
A.J.~Baltz, C.~Chasman, S.N.~White, Nucl. Instr. and Methods in Phys. Research A {\bf 417}, 1 (1998).

\bibitem{ALICE}
K. Aamodt et al. (ALICE Collaboration), J. Instrum. {\bf 3}, S08002 (2008).

\bibitem{Lepretre82}
A. Lepretre et al., Nucl. Phys. A {\bf 390}, 221 (1982).

\bibitem{Pshenichnov99}
I.A. Pshenichnov et al., Phys. Rev. C {\bf 60},  044901 (1999).

\bibitem{CERN}
M.B. Golubeva et al., Phys. Rev. C {\bf 71}, 024905 (2005).

\bibitem{RHIC}
M. Chiu et al., Phys. Rev. Lett. {\bf 89}, 012302 (2002).

\bibitem{vdM}
S. van der Meer, CERN/ISR-PO/68-31 (1968).

\bibitem{vdMA}
K.Oyama for the ALICE Collaboration, Proceedings of the LHC Lumi Days 2012, to be published.  

\bibitem{Alici}
A. Alici et al., CERN-ATS-Note-2011-016 PERF (LHC BCNWG Note 2).

\bibitem{Jeff}
A. Jeff, Proceedings of the LHC Lumi Days 2012, to be published.  

\bibitem{BCNWG}
A. Alici et al., CERN-ATS-Note-2012-029 PERF (LHC BCNWG Note 4).

\bibitem{ZDC}
R.~Arnaldi et al., Nucl.~Instr.~and~Meth. A {\bf 564}, 235 (2006); 
N.~De~Marco et al., IEEE Trans. on Nucl. Science {\bf 54} {\bf No.3}, 567 (2007).

\bibitem{Baltz}
A.J. Baltz et al., Physics Reports {\bf 458} (2008); A.J. Baltz, S.N. White, private communications.

\bibitem{higing}
X.-N. Wang and M. Gyulassy, Phys. Rev. D {\bf 44}, 3501 (1991).

\bibitem{TDR}
ALICE Zero Degree Calorimeter Technical Design Report, CERN/LHCC {\bf 99-5}, ALICE TDR 3 (1999), https://edms.cern.ch/document/398933/1.

\bibitem{Chadwick95}
M.B. Chadwick, P.G. Young, Acta Phys. Slov. {\bf 45}, 633 (1995).

\end{thebibliography}
\newenvironment{acknowledgement}{\relax}{\relax}
\begin{acknowledgement}
\section{Acknowledgements}
The ALICE collaboration would like to thank all its engineers and technicians for their invaluable contributions to the construction of the experiment and the CERN accelerator teams for the outstanding performance of the LHC complex.
\\
The ALICE collaboration acknowledges the following funding agencies for their support in building and
running the ALICE detector:
 \\
Calouste Gulbenkian Foundation from Lisbon and Swiss Fonds Kidagan, Armenia;
 \\
Conselho Nacional de Desenvolvimento Cient\'{\i}fico e Tecnol\'{o}gico (CNPq), Financiadora de Estudos e Projetos (FINEP),
Funda\c{c}\~{a}o de Amparo \`{a} Pesquisa do Estado de S\~{a}o Paulo (FAPESP);
 \\
National Natural Science Foundation of China (NSFC), the Chinese Ministry of Education (CMOE)
and the Ministry of Science and Technology of China (MSTC);
 \\
Ministry of Education and Youth of the Czech Republic;
 \\
Danish Natural Science Research Council, the Carlsberg Foundation and the Danish National Research Foundation;
 \\
The European Research Council under the European Community's Seventh Framework Programme;
 \\
Helsinki Institute of Physics and the Academy of Finland;
 \\
French CNRS-IN2P3, the `Region Pays de Loire', `Region Alsace', `Region Auvergne' and CEA, France;
 \\
German BMBF and the Helmholtz Association;
\\
General Secretariat for Research and Technology, Ministry of
Development, Greece;
\\
Hungarian OTKA and National Office for Research and Technology (NKTH);
 \\
Department of Atomic Energy and Department of Science and Technology of the Government of India;
 \\
Istituto Nazionale di Fisica Nucleare (INFN) of Italy;
 \\
MEXT Grant-in-Aid for Specially Promoted Research, Ja\-pan;
 \\
Joint Institute for Nuclear Research, Dubna;
 \\
National Research Foundation of Korea (NRF);
 \\
CONACYT, DGAPA, M\'{e}xico, ALFA-EC and the HELEN Program (High-Energy physics Latin-American--European Network);
 \\
Stichting voor Fundamenteel Onderzoek der Materie (FOM) and the Nederlandse Organisatie voor Wetenschappelijk Onderzoek (NWO), Netherlands;
 \\
Research Council of Norway (NFR);
 \\
Polish Ministry of Science and Higher Education;
 \\
National Authority for Scientific Research - NASR (Autoritatea Na\c{t}ional\u{a} pentru Cercetare \c{S}tiin\c{t}ific\u{a} - ANCS);
 \\
Federal Agency of Science of the Ministry of Education and Science of Russian Federation, International Science and
Technology Center, Russian Academy of Sciences, Russian Federal Agency of Atomic Energy, Russian Federal Agency for Science and Innovations and CERN-INTAS;
 \\
Ministry of Education of Slovakia;
 \\
Department of Science and Technology, South Africa;
 \\
CIEMAT, EELA, Ministerio de Educaci\'{o}n y Ciencia of Spain, Xunta de Galicia (Conseller\'{\i}a de Educaci\'{o}n),
CEA\-DEN, Cubaenerg\'{\i}a, Cuba, and IAEA (International Atomic Energy Agency);
 \\
Swedish Research Council (VR) and Knut $\&$ Alice Wallenberg
Foundation (KAW);
 \\
Ukraine Ministry of Education and Science;
 \\
United Kingdom Science and Technology Facilities Council (STFC);
 \\
The United States Department of Energy, the United States National
Science Foundation, the State of Texas, and the State of Ohio.

\end{acknowledgement}
\newpage
%
%
\appendix
\section{The ALICE Collaboration}
\label{app:collab}

\begingroup
\small
\begin{flushleft}
B.~Abelev\Irefn{org1234}\And
J.~Adam\Irefn{org1274}\And
D.~Adamov\'{a}\Irefn{org1283}\And
A.M.~Adare\Irefn{org1260}\And
M.M.~Aggarwal\Irefn{org1157}\And
G.~Aglieri~Rinella\Irefn{org1192}\And
A.G.~Agocs\Irefn{org1143}\And
A.~Agostinelli\Irefn{org1132}\And
S.~Aguilar~Salazar\Irefn{org1247}\And
Z.~Ahammed\Irefn{org1225}
% ...
\renewcommand\labelenumi{\textsuperscript{\theenumi}~}
\section*{Affiliation notes}
\renewcommand\theenumi{\roman{enumi}}
\begin{Authlist}
\item \Adef{M.V.Lomonosov Moscow State University, D.V.Skobeltsyn Institute of Nuclear Physics, Moscow, Russia}Also at: M.V.Lomonosov Moscow State University, D.V.Skobeltsyn Institute of Nuclear Physics, Moscow, Russia
\item \Adef{Institute of Nuclear Sciences, Belgrade, Serbia}Also at: "Vin\v{c}a" Institute of Nuclear Sciences, Belgrade, Serbia
\end{Authlist}
\section*{Collaboration Institutes}
\renewcommand\theenumi{\arabic{enumi}~}
\begin{Authlist}
\item \Idef{org1279}Benem\'{e}rita Universidad Aut\'{o}noma de Puebla, Puebla, Mexico
\item \Idef{org1220}Bogolyubov Institute for Theoretical Physics, Kiev, Ukraine
\item \Idef{org1262}Budker Institute for Nuclear Physics, Novosibirsk, Russia
\item \Idef{org1292}California Polytechnic State University, San Luis Obispo, California, United States
\item \Idef{org14939}Centre de Calcul de l'IN2P3, Villeurbanne, France
\item \Idef{org1197}Centro de Aplicaciones Tecnol\'{o}gicas y Desarrollo Nuclear (CEADEN), Havana, Cuba
\item \Idef{org1242}Centro de Investigaciones Energ\'{e}ticas Medioambientales y Tecnol\'{o}gicas (CIEMAT), Madrid, Spain
\item \Idef{org1244}Centro de Investigaci\'{o}n y de Estudios Avanzados (CINVESTAV), Mexico City and M\'{e}rida, Mexico
\item \Idef{org1335}Centro Fermi -- Centro Studi e Ricerche e Museo Storico della Fisica ``Enrico Fermi'', Rome, Italy
\item \Idef{org17347}Chicago State University, Chicago, United States
\item \Idef{org1288}Commissariat \`{a} l'Energie Atomique, IRFU, Saclay, France
\item \Idef{org1294}Departamento de F\'{\i}sica de Part\'{\i}culas and IGFAE, Universidad de Santiago de Compostela, Santiago de Compostela, Spain
\item \Idef{org1106}Department of Physics Aligarh Muslim University, Aligarh, India
\item \Idef{org1121}Department of Physics and Technology, University of Bergen, Bergen, Norway
\item \Idef{org1162}Department of Physics, Ohio State University, Columbus, Ohio, United States
\item \Idef{org1300}Department of Physics, Sejong University, Seoul, South Korea
\item \Idef{org1268}Department of Physics, University of Oslo, Oslo, Norway
\item \Idef{org1145}Dipartimento di Fisica dell'Universit\`{a} and Sezione INFN, Cagliari, Italy
\item \Idef{org1270}Dipartimento di Fisica dell'Universit\`{a} and Sezione INFN, Padova, Italy
\item \Idef{org1315}Dipartimento di Fisica dell'Universit\`{a} and Sezione INFN, Trieste, Italy
\item \Idef{org1132}Dipartimento di Fisica dell'Universit\`{a} and Sezione INFN, Bologna, Italy
\item \Idef{org1285}Dipartimento di Fisica dell'Universit\`{a} `La Sapienza' and Sezione INFN, Rome, Italy
\item \Idef{org1154}Dipartimento di Fisica e Astronomia dell'Universit\`{a} and Sezione INFN, Catania, Italy
\item \Idef{org1290}Dipartimento di Fisica `E.R.~Caianiello' dell'Universit\`{a} and Gruppo Collegato INFN, Salerno, Italy
\item \Idef{org1312}Dipartimento di Fisica Sperimentale dell'Universit\`{a} and Sezione INFN, Turin, Italy
\item \Idef{org1103}Dipartimento di Scienze e Tecnologie Avanzate dell'Universit\`{a} del Piemonte Orientale and Gruppo Collegato INFN, Alessandria, Italy
\item \Idef{org1114}Dipartimento Interateneo di Fisica `M.~Merlin' and Sezione INFN, Bari, Italy
\item \Idef{org1237}Division of Experimental High Energy Physics, University of Lund, Lund, Sweden
\item \Idef{org1192}European Organization for Nuclear Research (CERN), Geneva, Switzerland
\item \Idef{org1227}Fachhochschule K\"{o}ln, K\"{o}ln, Germany
\item \Idef{org1122}Faculty of Engineering, Bergen University College, Bergen, Norway
\item \Idef{org1136}Faculty of Mathematics, Physics and Informatics, Comenius University, Bratislava, Slovakia
\item \Idef{org1274}Faculty of Nuclear Sciences and Physical Engineering, Czech Technical University in Prague, Prague, Czech Republic
\item \Idef{org1229}Faculty of Science, P.J.~\v{S}af\'{a}rik University, Ko\v{s}ice, Slovakia
\item \Idef{org1184}Frankfurt Institute for Advanced Studies, Johann Wolfgang Goethe-Universit\"{a}t Frankfurt, Frankfurt, Germany
\item \Idef{org1215}Gangneung-Wonju National University, Gangneung, South Korea
\item \Idef{org1212}Helsinki Institute of Physics (HIP) and University of Jyv\"{a}skyl\"{a}, Jyv\"{a}skyl\"{a}, Finland
\item \Idef{org1203}Hiroshima University, Hiroshima, Japan
\item \Idef{org1329}Hua-Zhong Normal University, Wuhan, China
\item \Idef{org1254}Indian Institute of Technology, Mumbai, India
\item \Idef{org36378}Indian Institute of Technology Indore (IIT), Indore, India
\item \Idef{org1266}Institut de Physique Nucl\'{e}aire d'Orsay (IPNO), Universit\'{e} Paris-Sud, CNRS-IN2P3, Orsay, France
\item \Idef{org1277}Institute for High Energy Physics, Protvino, Russia
\item \Idef{org1249}Institute for Nuclear Research, Academy of Sciences, Moscow, Russia
\item \Idef{org1320}Nikhef, National Institute for Subatomic Physics and Institute for Subatomic Physics of Utrecht University, Utrecht, Netherlands
\item \Idef{org1250}Institute for Theoretical and Experimental Physics, Moscow, Russia
\item \Idef{org1230}Institute of Experimental Physics, Slovak Academy of Sciences, Ko\v{s}ice, Slovakia
\item \Idef{org1127}Institute of Physics, Bhubaneswar, India
\item \Idef{org1275}Institute of Physics, Academy of Sciences of the Czech Republic, Prague, Czech Republic
\item \Idef{org1139}Institute of Space Sciences (ISS), Bucharest, Romania
\item \Idef{org27399}Institut f\"{u}r Informatik, Johann Wolfgang Goethe-Universit\"{a}t Frankfurt, Frankfurt, Germany
\item \Idef{org1185}Institut f\"{u}r Kernphysik, Johann Wolfgang Goethe-Universit\"{a}t Frankfurt, Frankfurt, Germany
\item \Idef{org1177}Institut f\"{u}r Kernphysik, Technische Universit\"{a}t Darmstadt, Darmstadt, Germany
\item \Idef{org1256}Institut f\"{u}r Kernphysik, Westf\"{a}lische Wilhelms-Universit\"{a}t M\"{u}nster, M\"{u}nster, Germany
\item \Idef{org1246}Instituto de Ciencias Nucleares, Universidad Nacional Aut\'{o}noma de M\'{e}xico, Mexico City, Mexico
\item \Idef{org1247}Instituto de F\'{\i}sica, Universidad Nacional Aut\'{o}noma de M\'{e}xico, Mexico City, Mexico
\item \Idef{org23333}Institut of Theoretical Physics, University of Wroclaw
\item \Idef{org1308}Institut Pluridisciplinaire Hubert Curien (IPHC), Universit\'{e} de Strasbourg, CNRS-IN2P3, Strasbourg, France
\item \Idef{org1182}Joint Institute for Nuclear Research (JINR), Dubna, Russia
\item \Idef{org1143}KFKI Research Institute for Particle and Nuclear Physics, Hungarian Academy of Sciences, Budapest, Hungary
\item \Idef{org1199}Kirchhoff-Institut f\"{u}r Physik, Ruprecht-Karls-Universit\"{a}t Heidelberg, Heidelberg, Germany
\item \Idef{org20954}Korea Institute of Science and Technology Information, Daejeon, South Korea
\item \Idef{org1160}Laboratoire de Physique Corpusculaire (LPC), Clermont Universit\'{e}, Universit\'{e} Blaise Pascal, CNRS--IN2P3, Clermont-Ferrand, France
\item \Idef{org1194}Laboratoire de Physique Subatomique et de Cosmologie (LPSC), Universit\'{e} Joseph Fourier, CNRS-IN2P3, Institut Polytechnique de Grenoble, Grenoble, France
\item \Idef{org1187}Laboratori Nazionali di Frascati, INFN, Frascati, Italy
\item \Idef{org1232}Laboratori Nazionali di Legnaro, INFN, Legnaro, Italy
\item \Idef{org1125}Lawrence Berkeley National Laboratory, Berkeley, California, United States
\item \Idef{org1234}Lawrence Livermore National Laboratory, Livermore, California, United States
\item \Idef{org1251}Moscow Engineering Physics Institute, Moscow, Russia
\item \Idef{org1140}National Institute for Physics and Nuclear Engineering, Bucharest, Romania
\item \Idef{org1165}Niels Bohr Institute, University of Copenhagen, Copenhagen, Denmark
\item \Idef{org1109}Nikhef, National Institute for Subatomic Physics, Amsterdam, Netherlands
\item \Idef{org1283}Nuclear Physics Institute, Academy of Sciences of the Czech Republic, \v{R}e\v{z} u Prahy, Czech Republic
\item \Idef{org1264}Oak Ridge National Laboratory, Oak Ridge, Tennessee, United States
\item \Idef{org1189}Petersburg Nuclear Physics Institute, Gatchina, Russia
\item \Idef{org1170}Physics Department, Creighton University, Omaha, Nebraska, United States
\item \Idef{org1157}Physics Department, Panjab University, Chandigarh, India
\item \Idef{org1112}Physics Department, University of Athens, Athens, Greece
\item \Idef{org1152}Physics Department, University of Cape Town, iThemba LABS, Cape Town, South Africa
\item \Idef{org1209}Physics Department, University of Jammu, Jammu, India
\item \Idef{org1207}Physics Department, University of Rajasthan, Jaipur, India
\item \Idef{org1200}Physikalisches Institut, Ruprecht-Karls-Universit\"{a}t Heidelberg, Heidelberg, Germany
\item \Idef{org1325}Purdue University, West Lafayette, Indiana, United States
\item \Idef{org1281}Pusan National University, Pusan, South Korea
\item \Idef{org1176}Research Division and ExtreMe Matter Institute EMMI, GSI Helmholtzzentrum f\"ur Schwerionenforschung, Darmstadt, Germany
\item \Idef{org1334}Rudjer Bo\v{s}kovi\'{c} Institute, Zagreb, Croatia
\item \Idef{org1298}Russian Federal Nuclear Center (VNIIEF), Sarov, Russia
\item \Idef{org1252}Russian Research Centre Kurchatov Institute, Moscow, Russia
\item \Idef{org1224}Saha Institute of Nuclear Physics, Kolkata, India
\item \Idef{org1130}School of Physics and Astronomy, University of Birmingham, Birmingham, United Kingdom
\item \Idef{org1338}Secci\'{o}n F\'{\i}sica, Departamento de Ciencias, Pontificia Universidad Cat\'{o}lica del Per\'{u}, Lima, Peru
\item \Idef{org1316}Sezione INFN, Trieste, Italy
\item \Idef{org1271}Sezione INFN, Padova, Italy
\item \Idef{org1313}Sezione INFN, Turin, Italy
\item \Idef{org1286}Sezione INFN, Rome, Italy
\item \Idef{org1146}Sezione INFN, Cagliari, Italy
\item \Idef{org1133}Sezione INFN, Bologna, Italy
\item \Idef{org1115}Sezione INFN, Bari, Italy
\item \Idef{org1155}Sezione INFN, Catania, Italy
\item \Idef{org1322}Soltan Institute for Nuclear Studies, Warsaw, Poland
\item \Idef{org36377}Nuclear Physics Group, STFC Daresbury Laboratory, Daresbury, United Kingdom
\item \Idef{org1258}SUBATECH, Ecole des Mines de Nantes, Universit\'{e} de Nantes, CNRS-IN2P3, Nantes, France
\item \Idef{org1304}Technical University of Split FESB, Split, Croatia
\item \Idef{org1168}The Henryk Niewodniczanski Institute of Nuclear Physics, Polish Academy of Sciences, Cracow, Poland
\item \Idef{org17361}The University of Texas at Austin, Physics Department, Austin, TX, United States
\item \Idef{org1173}Universidad Aut\'{o}noma de Sinaloa, Culiac\'{a}n, Mexico
\item \Idef{org1296}Universidade de S\~{a}o Paulo (USP), S\~{a}o Paulo, Brazil
\item \Idef{org1149}Universidade Estadual de Campinas (UNICAMP), Campinas, Brazil
\item \Idef{org1239}Universit\'{e} de Lyon, Universit\'{e} Lyon 1, CNRS/IN2P3, IPN-Lyon, Villeurbanne, France
\item \Idef{org1205}University of Houston, Houston, Texas, United States
\item \Idef{org20371}University of Technology and Austrian Academy of Sciences, Vienna, Austria
\item \Idef{org1222}University of Tennessee, Knoxville, Tennessee, United States
\item \Idef{org1310}University of Tokyo, Tokyo, Japan
\item \Idef{org1318}University of Tsukuba, Tsukuba, Japan
\item \Idef{org21360}Eberhard Karls Universit\"{a}t T\"{u}bingen, T\"{u}bingen, Germany
\item \Idef{org1225}Variable Energy Cyclotron Centre, Kolkata, India
\item \Idef{org1306}V.~Fock Institute for Physics, St. Petersburg State University, St. Petersburg, Russia
\item \Idef{org1323}Warsaw University of Technology, Warsaw, Poland
\item \Idef{org1179}Wayne State University, Detroit, Michigan, United States
\item \Idef{org1260}Yale University, New Haven, Connecticut, United States
\item \Idef{org1332}Yerevan Physics Institute, Yerevan, Armenia
\item \Idef{org15649}Yildiz Technical University, Istanbul, Turkey
\item \Idef{org1301}Yonsei University, Seoul, South Korea
\item \Idef{org1327}Zentrum f\"{u}r Technologietransfer und Telekommunikation (ZTT), Fachhochschule Worms, Worms, Germany
\end{Authlist}
\endgroup

\end{document}